\documentstyle[twocolumn,aps,prb,epsfig]{revtex}
\begin{document}
\draft
\preprint{}

\wideabs{
\title{
 Anisotropic Antiferromagnetic Spin Chains in a Transverse Field:\\
Reentrant Behavior of the Staggered Magnetization}
\author{ Yasuhiro Hieida, Kouichi Okunishi$^{2}$ and Yasuhiro Akutsu$^{3}$ }
\address{
Department of Physics, Faculty of Science, Kobe University, Kobe 657-8501, Japan.\\  
$^{2}_{~}$  Department of Physics, Faculty of Science, Niigata University,  Igarashi 2, Niigata 950-2181, Japan\\
$^{3}_{~}$Department of Physics, Graduate School of Science, Osaka
University, Toyonaka, Osaka 560-0043, Japan. 
}
\date{\today}
\maketitle

\begin{abstract}
We investigate one-dimensional $S=1/2$ and $S=1$ antiferromagnetic
quantum spin chains with easy-axis anisotropies in a transverse field.
We calculate both the uniform magnetization and the staggered
magnetization, using a variant of the density matrix renormalization
group method.  We find that the staggered magnetizations exhibit 
reentrant behavior as functions of the transverse field, where the
competition between the classical N{\' e}el order and the quantum
fluctuations plays an important role.
We also discuss the critical behavior associated with the staggered
magnetizations.
\end{abstract}

\pacs{75.10.Jm, 75.30.Gw, 75.40.Cx}
}

\section{Introduction}

Recently, magnetization processes of one-dimensional (1D) quantum spin systems have been attracting much interest from theoretical and experimental view points.
Particularly, various types of field-induced critical behavior in the magnetization process, such as a square-root behavior associated with the excitation gap~\cite{Tsvelik,Affleck,square-root-behavior}, the magnetization plateau~\cite{plateau-papers}, the cusp singularity~\cite{Cusp}, etc. have been investigated intensively. 
In most of these theoretical analyses of the magnetization process, the external field is supposed to be applied in the longitudinal direction.
For the case of a system having an anisotropy, however, magnetic properties of the system may be essentially different, depending on whether the external
field is applied in the traverse or longitudinal directions.
In this paper, we wish to make {\it a quantitative analysis} for the magnetic properties of antiferromagnetic quantum spin chains with  easy-axis anisotropies in a transverse field.

In analyzing an anisotropic model in the transverse magnetic field, an essential point is that the ``total $S^z$'' of the system is not a conservative quantity, since the symmetry of the model is reduced to be $Z_2$.
Then, the property of the system is characterized with the staggered magnetization.
For example, the appearance of the staggered magnetization is known for the $S=1/2$ XXZ spin chain in the Ising limit, while the isotropic XXX model has no N{\' e}el order; 
The anisotropy favors the classical N{\' e}el order, whereas the transverse field term induces a quantum fluctuation in addition to the usual XY-term.
What we are interested in here is how the system behaves between the Ising limit and the isotropic limit.
We discuss the physics developed by the competition between the anisotropy and the quantum fluctuations, based on quantitative calculations of the staggered magnetization.

For the purpose of computing the magnetization curve, we employ the product wave function renormalization group(PWFRG) method,\cite{NO,HOA} which is a variant of the density matrix renormalization group(DMRG) method\cite{orig-DMRG1}.
The PWFRG is successfully applied to various types of the magnetization process,\cite{HOA,SA,square-root-behavior,Cusp,S-ISSP}
and further it can treat a system without the total-$S^z$ conservation efficiently.
We calculate the staggered magnetization in the easy-axis direction and the uniform magnetization in the transverse one, for the infinite-length chains.
Here it should be noted that the lack of the total-$S^z$ conservation makes the exact diagonalization study difficult for the anisotropic models in the transverse field.

For the $S=1/2$ XXZ chain, we find the reentrant behavior of the staggered magnetization,  which is partially supported by the exact results at some special parameters.
We further discuss the $S=1$ XXZ spin chain and the $S=1$ Heisenberg spin chain with the single-ion anisotropy, where the existence of the Haldane gap makes the physics of the systems more fruitful.\cite{Tsvelik,Affleck,Tasaki,SaTa,Mikeska}
We verify that the phase transitions occur twice at the lower critical field and the higher critical one; The former field is associated with the Haldane gap, and the later one can be connected with the saturation field of the isotropic case adiabatically. 
We also discuss the critical behavior characterized by the staggered magnetizations.

This paper is organized as follows: In \S\ref{sec:half-spin}, we describe the $S=1/2$ XXZ spin chain in a transverse field.
In \S\ref{sec:spin-1}, we consider the $S=1$ quantum spin chains with the
XXZ-type anisotropy(\S\ref{sec:spin-1}-A) and the single-ion anisotropy(\S\ref{sec:spin-1}-B).
The last section is devoted to the summary.

\section{The $S=1/2$ case}
\label{sec:half-spin}

The $S=1/2$ XXZ chain is one of the most fundamental models among spin chains of $S=1/2$. In this section, we consider the XXZ chain  in a transverse field. We write the Hamiltonian ${\cal H}$ as 
\begin{equation}
\label{halfxxz}
{\cal H}=\sum_{i} [ J_{\parallel} ( S^x_i S^x_{i+1} +S^y_i S^y_{i+1} )
                + J_\perp S^z_i S^z_{i+1}]
        -\Gamma \sum_{i} S^x_i,
\end{equation}
where $S^\alpha_i$ is the $\alpha$ component of the $S=1/2$ spin operator at $i$-th site, and $\Gamma$ is the strength of the transverse field. 
The exchange coupling constant in easy-plane($xy$-plane) is denoted by $J_{\parallel}$ and that of the easy-axis($z$-axis) direction by $J_\perp$. 
We have set $g\mu_{\rm B}=1$ ($g$: $g$-factor, $\mu_{B}$: Bohr magneton) for simplicity.
In this section, we discuss the system (\ref{halfxxz}) in the range $0\le J_{\parallel}\le 1$ with fixing $J_\perp=1$.

The magnetic property of the Hamiltonian (\ref{halfxxz}) has been studied for several years. \cite{disorder-point,Existence-Neel,GMLH,Harada-1,Harada-2}. 
Particularly,  it should be remarked that there is a special value of the transverse field called a disorder point,  where the ground state is  written down analytically.~\cite{disorder-point,Existence-Neel}
In addition, Mori {\it et al.} have discussed how the excitation structure of eq. (\ref{halfxxz}) is affected by the transverse field.\cite{Harada-1,Harada-2}
However, the full magnetization process for general parameters of eq.(\ref{halfxxz}) has not been calculated quantitatively, except for the exactly solved cases $J_\parallel=0$ (the Ising model in a transverse field) and $J_\parallel=1$ (the Heisenberg model).

In the following, we calculate the uniform magnetization in $x$-direction $M_x \equiv \langle S^x_i \rangle $ and the staggered magnetization in $z$-direction $M_{\rm st} \equiv |\langle S^z_i \rangle |$, by using the PWFRG method.
Here we note that it is useful to consider the $\pi$-rotation of spins at every two sites(for example, at every even sites):
\begin{equation}
U \equiv \prod_{k\in\{\rm even\, sites\}} \exp[-i\pi S^x_k],
\end{equation}
since we can calculate the staggered magnetization as the uniform magnetization in the transformed system ${\cal H}'=U^\dagger {\cal H}U$. 
We have confirmed that the computed results for the transformed model agree with the staggered magnetization calculated directly for eq. (\ref{halfxxz}).

\begin{figure}
 \begin{center}
  \leavevmode  \epsfxsize=65mm
  \epsfbox{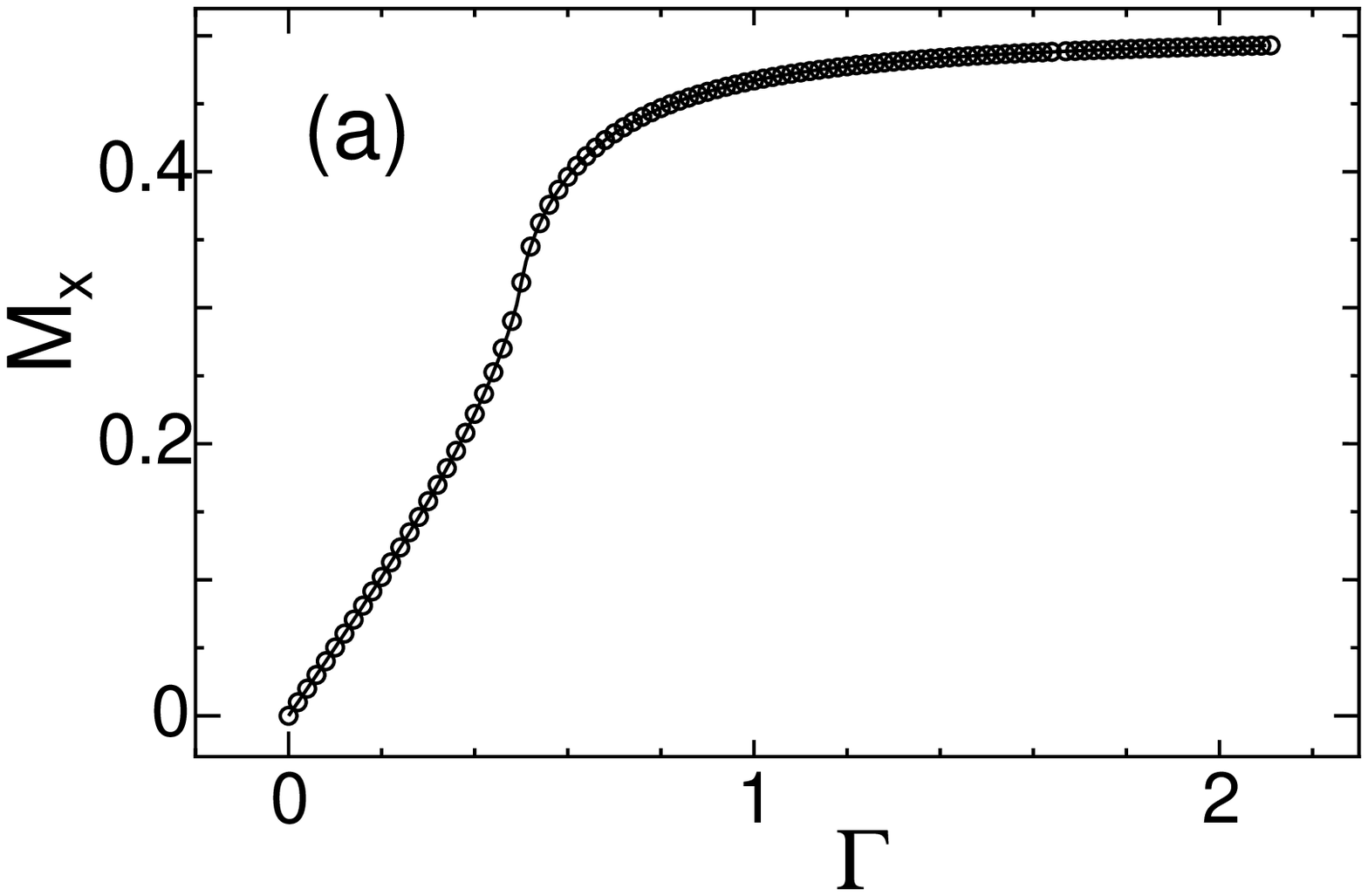}\\
  \leavevmode  \epsfxsize=65mm
  \epsfbox{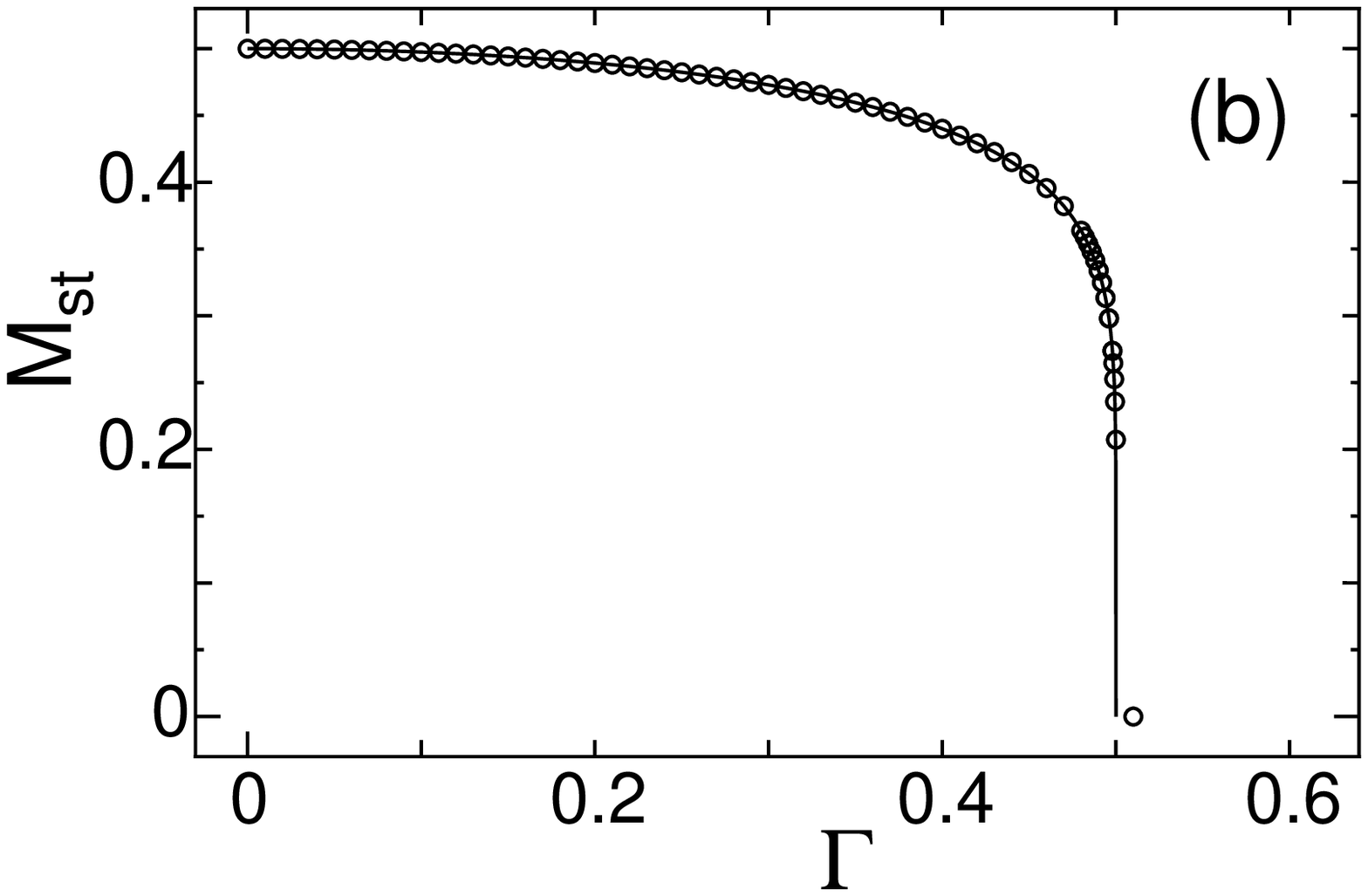}
  \end{center}
 \caption{
The PWFRG results(represented by $\circ$) for the Ising model in a transverse field:
 (a) the uniform magnetization $M_x$ and (b) the staggered magnetization $M_{\rm st}$. The exact solutions of $M_x$ and $M_{\rm st}$ are denoted by the solid lines.}
 \label{figITF}
\end{figure}

Before proceeding to the detailed analysis of the XXZ chain, we consider the Ising model in a transverse field (ITF) ($J_\parallel=0$), for which the analytical solution is known as follows~\cite{Pfeuty}:
\begin{equation}
M_x=\int_0^\pi \frac{dk}{2\pi}
\frac{(\cos{k}+2\Gamma)}{\sqrt{1+4\Gamma\cos{k}+4 \Gamma^2}}, 
\label{eq:ITF-Sx}
\end{equation}
and,
\begin{equation}
M_{\rm st} = \frac{1}{2}(1-4\Gamma^2)^{1/8}.
\label{eq:ITF-Sz}
\end{equation}
We can then check the efficiency of the PWFRG method, by comparing the obtained results with the above exact expressions.
In Fig.~\ref{figITF}, we can see that the computed results are in good agreement with the exact ones, where the PWFRG calculation is performed with the  number of retained bases $m=14$.  
Thus, we can expect that the PWFRG method with a relatively small $m$ yields accurate results in the following analysis of the XXZ spin chain.

\begin{figure}
 \begin{center}
  \leavevmode  \epsfxsize=65mm
  \epsfbox{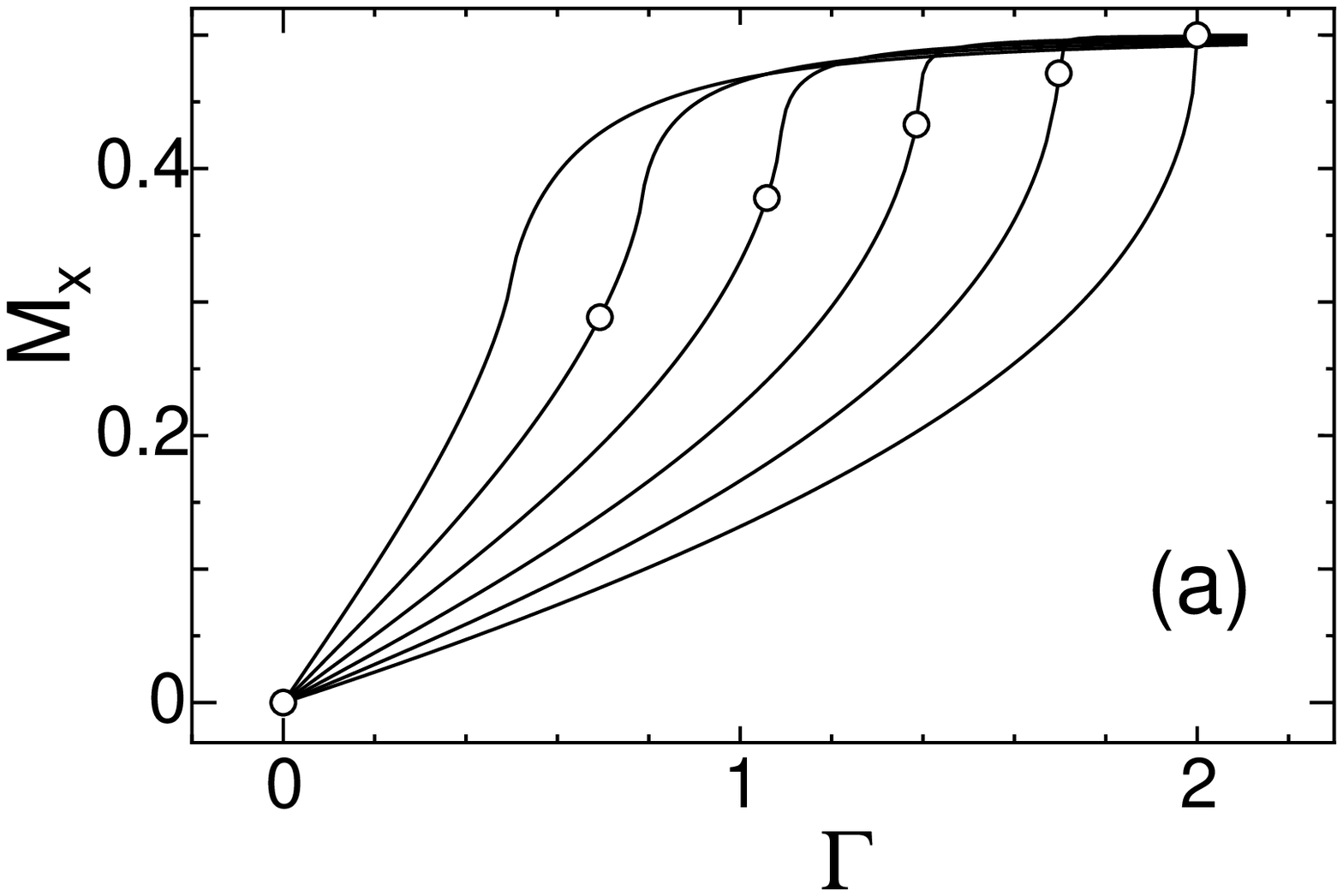}\\
  \leavevmode  \epsfxsize=65mm
  \epsfbox{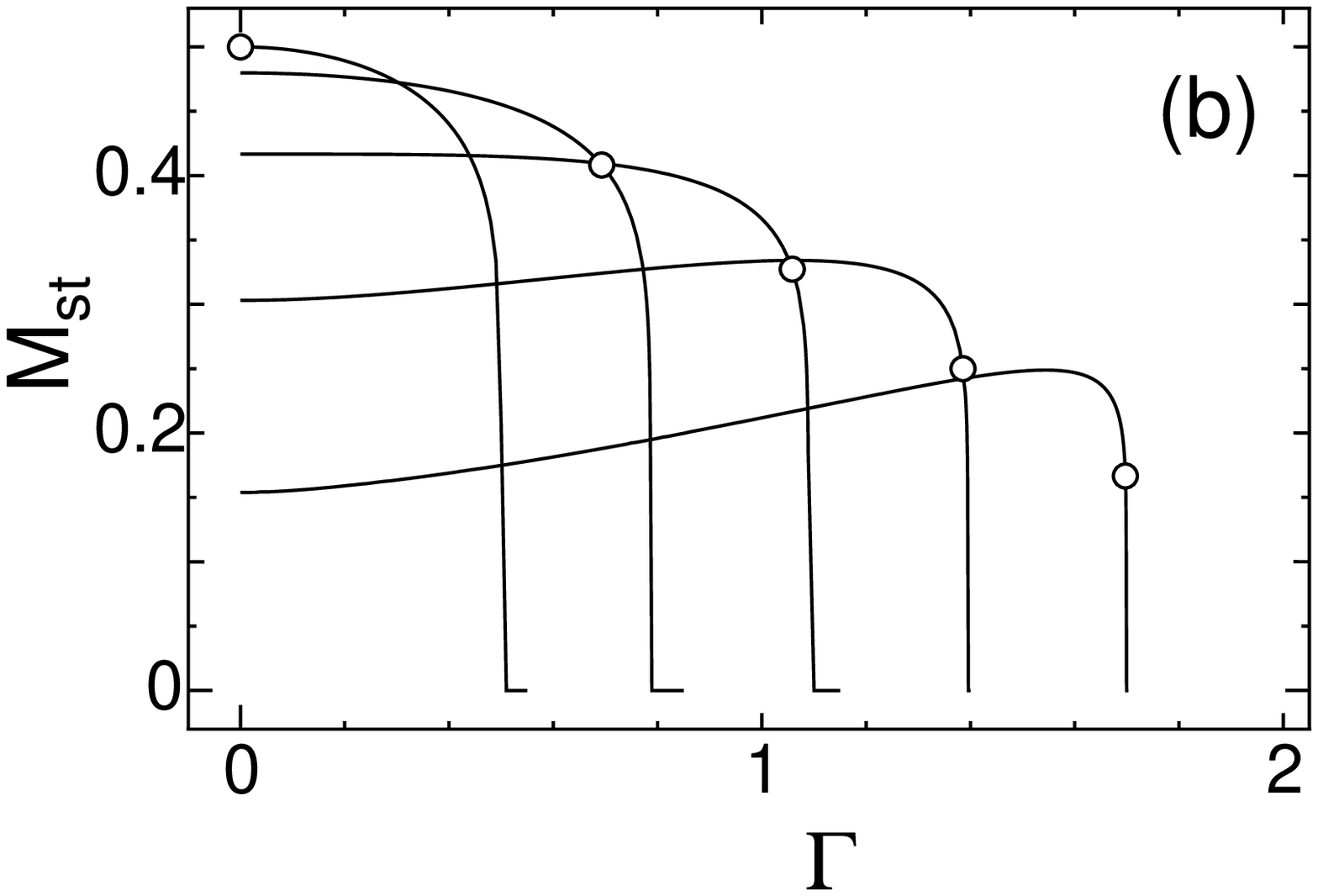}
 \end{center}
\caption{
The PWFRG results for the $S=1/2$ XXZ spin chain with various
$J_\parallel$: (a) the uniform magnetization $M_x$ and (b) the
staggered magnetization $M_{\rm st}$. 
The $\Gamma$-$M_x$ curves correspond to $J_\parallel=0$, $0.2$, $0.4$, $0.6$, $0.8$, $1$ from left to right.
The $\Gamma$-$M_{\rm st}$ curves correspond to $J_\parallel=0$, $0.2$, $0.4$, $0.6$, $0.8$, from left to right.
The symbol $\circ$ indicates the disorder point.}
\label{fighalfxxz}
\end{figure}

In Fig.~\ref{fighalfxxz}, we display the PWFRG results of the magnetization in the $x$ direction  $M_x$ for $J_\parallel=0$, $0.2$, $0.4$, $0.6$, $0.8$, $1$ and the staggered magnetization $M_{\rm st}$ for $J_\parallel=0$, $0.2$, $0.4$, $0.6$, $0.8$.
In Fig.\ref{fighalfxxz}-(a), the $M_x$ grows monotonously as the field $\Gamma$ is increased,  whereas in Fig.\ref{fighalfxxz}-(b), the  staggered magnetizations $M_{\rm st}$ for $J_\parallel > 0.4$ exhibit the reentrant behavior.
This reentrant behavior can be explained qualitatively in terms of the competition between the classical N{\' e}el order and two kinds of the quantum fluctuations.
In the low-field region, the system is under influence of the quantum fluctuation from the XY-term($J_{\parallel} \sum S^x_i S^x_{i+1} +S^y_i S^y_{i+1} $), while, in the high-field region, the effect of the transverse-field  becomes dominant.
Although both of them can decrease the classical N{\'e}el order, the
staggered magnetization is rather enhanced in the middle-field region.
This enhancement implies that the two quantum fluctuations conflict with each
other rather than reduce the N{\'e}el order. This picture is
consistent with the shape changes of the static structure factor in
the transverse field, as was reported in Ref. \cite{Harada-2}. 

Let us now proceed to analysis of details in the magnetization curve.
In the zero transverse field limit,  the staggered magnetization is calculated exactly:~\cite{Baxter-1,Kyoto-1}  
\begin{equation}
\left .M_{\rm st}\right |_{\Gamma=0} = \frac{1}{2}   \prod_{k=1}^{\infty}
    \left (  \frac{1-q^{2k}}{1+q^{2k}}  \right )^2 ,
\label{eq:ExactStagMag}
\end{equation}
where $q = 1/J_\parallel - \sqrt{1/J_\parallel^2 - 1 }$.
By comparing  the PWFRG results at $\Gamma=0$ with the exact values of eq.(\ref{eq:ExactStagMag}), we have confirmed that the obtained data for $J_\parallel\le 0.6$ agree with the exact values in 5  digits,  within the numbers of the retained bases $m=88$.

There is another special value of a transverse field  called disorder point, where the ground state can be calculated analytically.\cite{disorder-point,Existence-Neel}
The disorder-point field $\Gamma_{\rm dp}$ is given by
\begin{equation}
\Gamma_{\rm dp} = \sqrt{2 J_\parallel (J_\parallel+1)},
\end{equation}
and the exact values of the magnetizations are  obtained as 
\begin{equation}
M_x = \frac{1}{2}\sqrt{\frac{2 J_\parallel}{J_\parallel+1}},
\end{equation}
and
\begin{equation}
M_{\rm st} = \frac{1}{2}\sqrt{\frac{1-J_\parallel}{1+J_\parallel}}.
\label{eq:half-spin-disorder-Mst}
\end{equation}
In Figs~\ref{fighalfxxz}-(a) and (b) we have plotted the disorder point with the open circles, which are consistent with the curves calculated by the PWFRG.

Here we remark that we can obtain an inequality:
\begin{equation}
\left .M_{\rm st}\right |_{\Gamma=0}< \left .M_{\rm st}\right
|_{\Gamma=\Gamma_{\rm dp}},
\label{eq:half-spin-rel-1}
\end{equation}
for $J_\parallel > 0.73 $, by comparing eq.(\ref{eq:ExactStagMag}) and eq.(\ref{eq:half-spin-disorder-Mst}).
This inequality (\ref{eq:half-spin-rel-1}) guarantees the reentrant behavior of the staggered magnetization analytically in the range $J_\parallel >0.73$.
A precise numerical calculation of $M_{\rm st}$(using maximum $m=30$) illustrates that the reentrant behavior emerges for $J_\parallel\ge0.3984$.

\begin{figure}
 \begin{center}
  \leavevmode  \epsfxsize=65mm
  \epsfbox{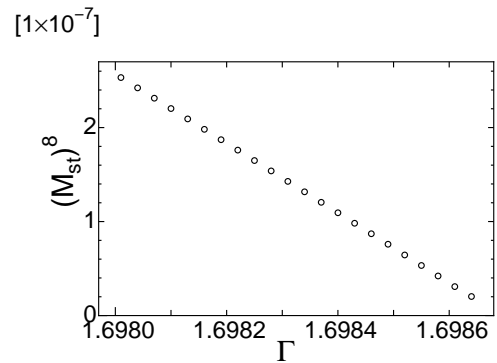}
 \end{center}
\caption{
The check of linearity of the $\Gamma$-$(M_{\rm st})^8$ curve in the case of $J_\parallel=0.8$ near the critical field. The good linearity(see eq.(\ref{eq:beta})) can be seen.
}
\label{halfxxzbeta}
\end{figure}

We next discuss the critical behavior of eq. (\ref{halfxxz}).
The critical field $\Gamma_s$ is defined by the field at which the staggered magnetization vanishes. 
Since the symmetry of the system is $Z_2$, the universality class is expected to be the two-dimensional(2D) classical Ising one, where the staggered magnetization behaves as 
\begin{equation}
M_{\rm st} \sim (\Gamma_{\rm s}-\Gamma)^{\beta}  \quad {\rm for} \quad \Gamma < \Gamma_{\rm s},
\label{eq:beta}
\end{equation}
with $\beta=1/8$.
For the case of $J_\parallel=0$,  we can see $\beta=1/8$ from the exact solution (\ref{eq:ITF-Sz}) easily.
For $0<J_\parallel <1$, we have checked the eq.(\ref{eq:beta}) with the PWFRG result of the staggered magnetization.
For example , we show the plot of $\Gamma$-$(M_{\rm st})^8$ for  $J_\parallel = 0.8$ in Fig.\ref{halfxxzbeta}, where we can see the good linearity of the plotted data.
We have performed the fitting of $M_{\rm st} = A (\Gamma_{\rm s}-\Gamma)^{\beta}$ with  fitting parameters $A$, $\beta$ and $\Gamma_{\rm s}$, and then obtain the exponent $\beta=0.124, 0.124$ and $0.125$, for $J_\parallel = 0.4, 0.6$ and $0.8$, respectively.
These values  are consistent with $\beta=1/8$.
Thus, we have verified that the universality class of the XXZ model at the critical transverse field is the 2D Ising one.

\section{ $S=1$ cases}
\label{sec:spin-1}

In this section, we consider the $S=1$ XXZ spin chain and the $S=1$
Heisenberg spin chain with a single-ion anisotropy. 
In contrast to the $S=1/2$ XXZ model,  the $S=1$ chains with small anisotropies have the Haldane gap\cite{Tsvelik,Affleck,Tasaki,Mikeska}. 
Then it is an interesting problem to clarify how the Haldane gap influences the competition between the classical N{\' e}el order and the transverse field.

\subsection{the $S=1$ XXZ model}

In this subsection, we consider the $S=1$  XXZ spin chain, whose Hamiltonian is written as 
\begin{equation}
{\cal H} =
\sum_{i} [
           J_{\parallel} ( S^x_i S^x_{i+1} +S^y_i  S^y_{i+1} )
         + J_\perp S^z_i S^z_{i+1}]
          -\Gamma \sum_{i} S^x_i,
\label{eqs1xxz}
\end{equation}
where $S^\alpha_i$ is the $\alpha$ component of the $S=1$ spin operator at $i$-th site. 
In this section, we fix $J_\parallel=1$ and vary $J_\perp$ in $J_\perp\ge 1$, for the later convenience.

\begin{figure}
 \begin{center}
  \leavevmode  \epsfxsize=65mm
  \epsfbox{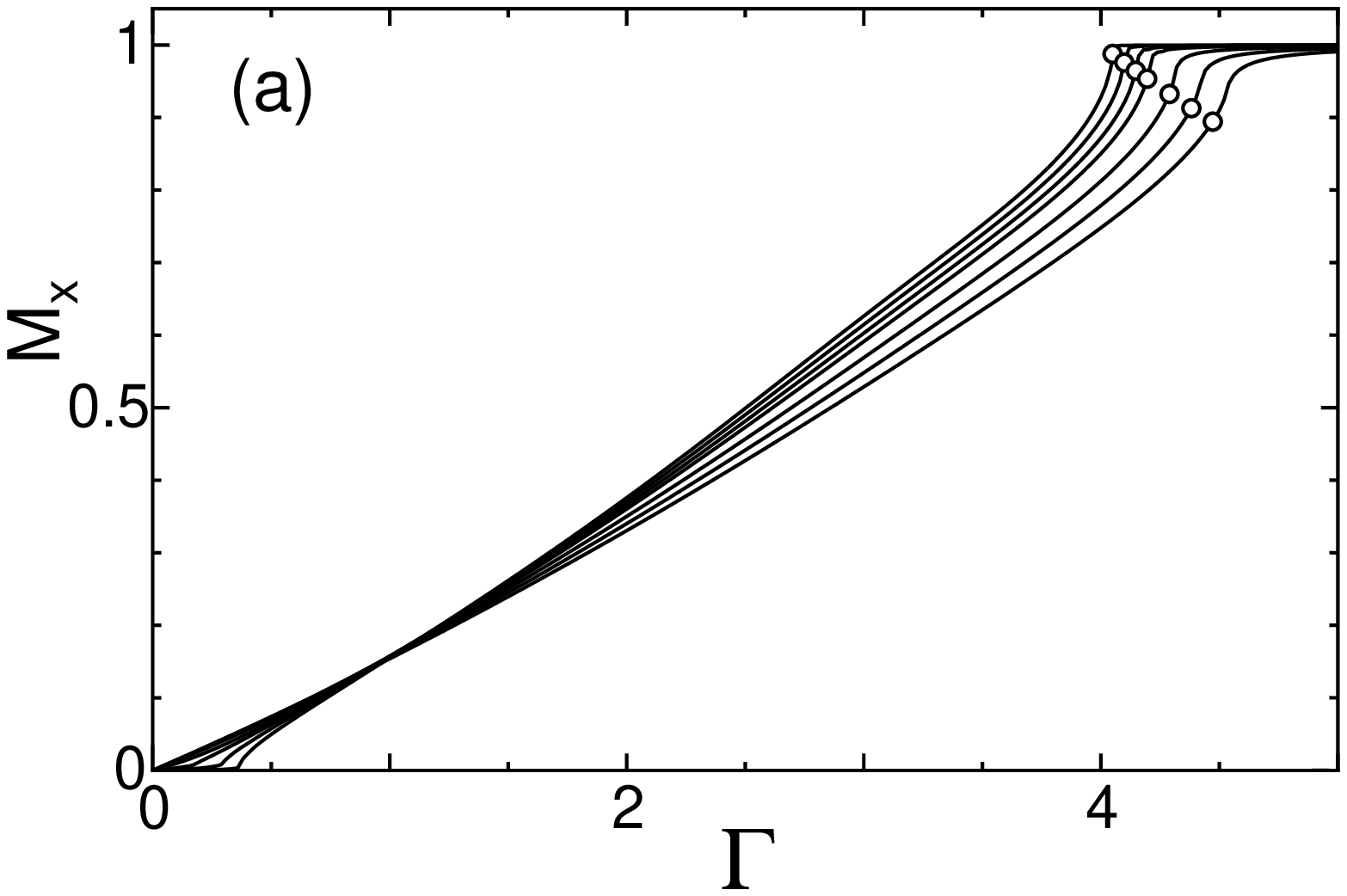}\\
  \leavevmode  \epsfxsize=65mm
  \epsfbox{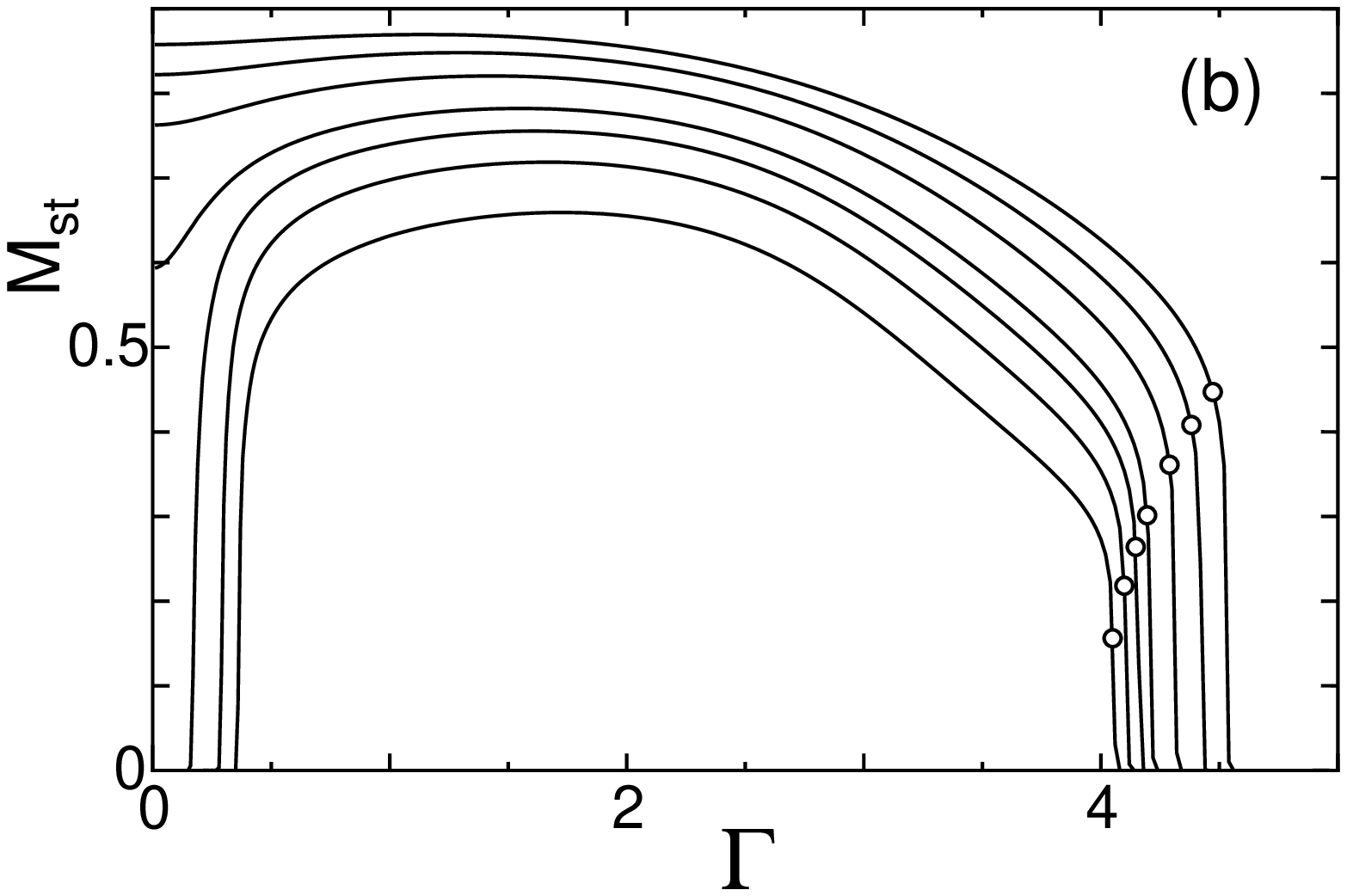}
 \end{center}
\caption{The PWFRG results for the $S=1$ XXZ spin chain: (a) the uniform magnetization $M_x$ and (b) the staggered magnetization $M_{\rm st}$. 
Each curve corresponds to $J_\perp=1.5$, $1.4$, $1.3$, $1.2$, $1.15$, $1.1$,
and  $1.05$
from top to bottom.
The open circle represents the disorder point for each $J_\perp$.}
\label{fs1xxz}
\end{figure}

In Fig.\ref{fs1xxz},  we show the staggered magnetization $M_{\rm st}$ and the uniform magnetization $M_x$.
The maximum number of the retained bases in the DMRG calculation is $m=70$.
In the figures, we also plot the disorder points for the $S=1$ XXZ model,\cite{disorder-point,Existence-Neel} which is given by 
\begin{eqnarray}
M_x = \sqrt{\frac{2}{1+J_\perp}}
\quad {\rm and} \quad 
M_{\rm st} = \sqrt{\frac{J_\perp-1}{J_\perp+1}},
\end{eqnarray}
with the disorder-point field $\Gamma_{\rm dp}=2\sqrt{2(1+J_{\perp})}$. 
We can see that they are in good agreement with the DMRG results. 
When the anisotropy is not so big($J_\perp <
1.18$),\cite{Haldane-Neel-1} the Haldane state is maintained up to the
critical value of the transverse field $\Gamma_{c1}$, where the phase
transition occurs accompanying the spontaneous staggered order(Fig.4-(b)).  
As $\Gamma$ is increased above $\Gamma_{c1}$, the staggered
magnetization grows rapidly.  
However, as $\Gamma$ is further increased, $M_{\rm st}$ turns to decrease and finally vanishes at the higher critical field $\Gamma_{c2}$.
In Fig. \ref{fs1xxz}-(a),  we show the uniform magnetization $M_x$,
where the anisotropy effect seems not to be significant. 
However, a notable point for this uniform magnetization is that $M_x$ has a small but finite value below the critical field $\Gamma_{c1}$, in contrast to the isotropic case where $M_x$ is exactly zero below $\Gamma_{c1}$ . 
In a real experiment, it may be important to distinguish such a behavior of $M_x$ from the finite temperature effect.

As $J_\perp$ is increased, the lower-critical field $\Gamma_{c1}$ becomes small and thus the region of the Haldane phase shrinks(see Fig.\ref{fs1xxz}-(a)). 
When $J_\perp$ is increased above the critical value of the anisotropy  $J_{\perp c}\simeq 1.18$, the spontaneous staggered magnetization appears at the zero transverse field,\cite{Haldane-Neel-1} and $M_{\rm st}\sim (J_\perp-J_{\perp c})^{1/8}$ is observed.\cite{Nomuramc}

For $J_{\perp} > J_{\perp c}$ the staggered magnetization exhibits the week reentrant behavior, which is similar to the $S=1/2$ case.
In the $J_\perp\to \infty$ limit, the system becomes to the $S=1$ version of the Ising model in a transverse field, without the reentrance of $M_{\rm st}$.

\begin{figure}
 \begin{center}
  \leavevmode  \epsfxsize=65mm
  \epsfbox{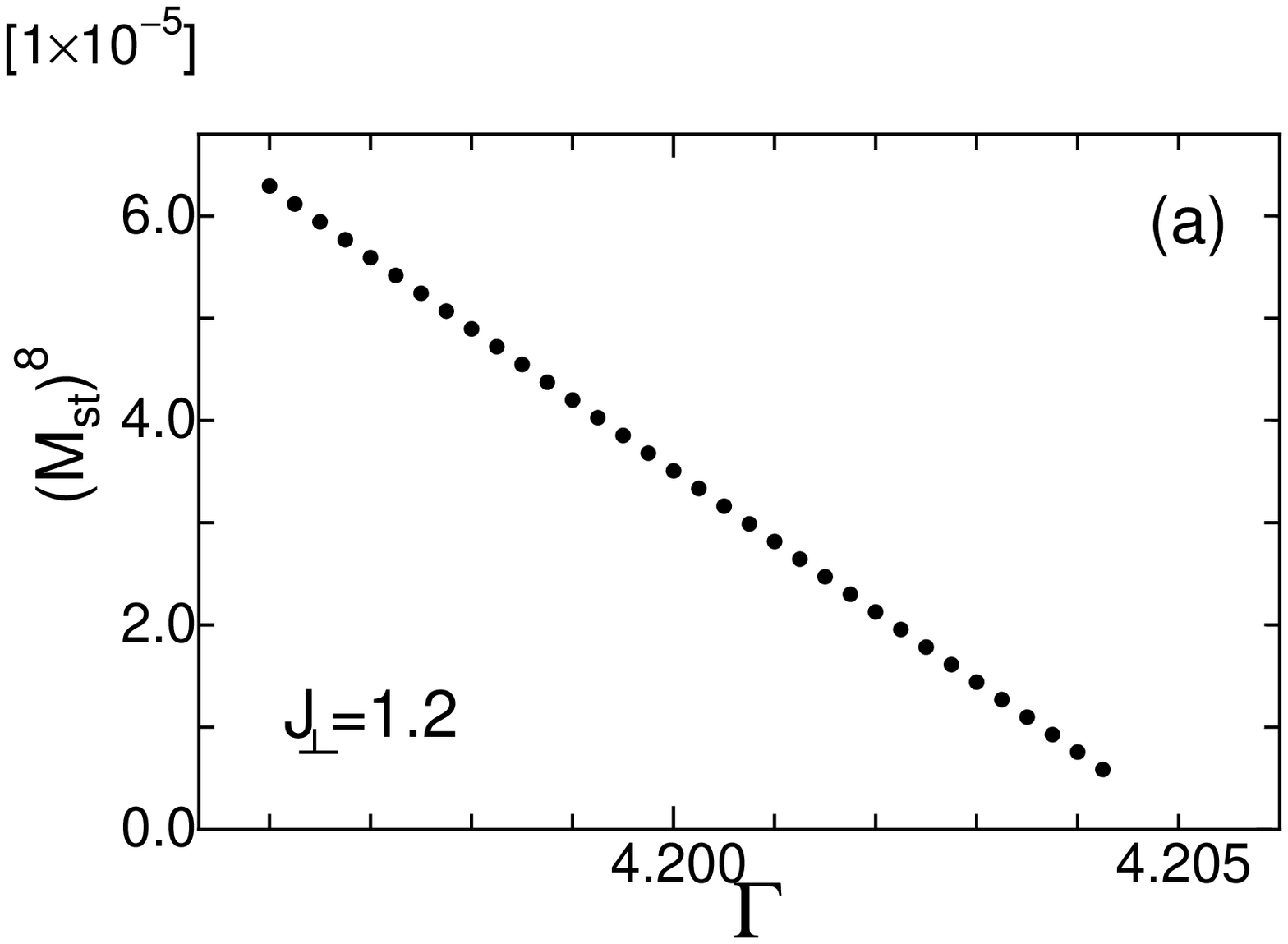}\\
  \leavevmode  \epsfxsize=65mm
  \epsfbox{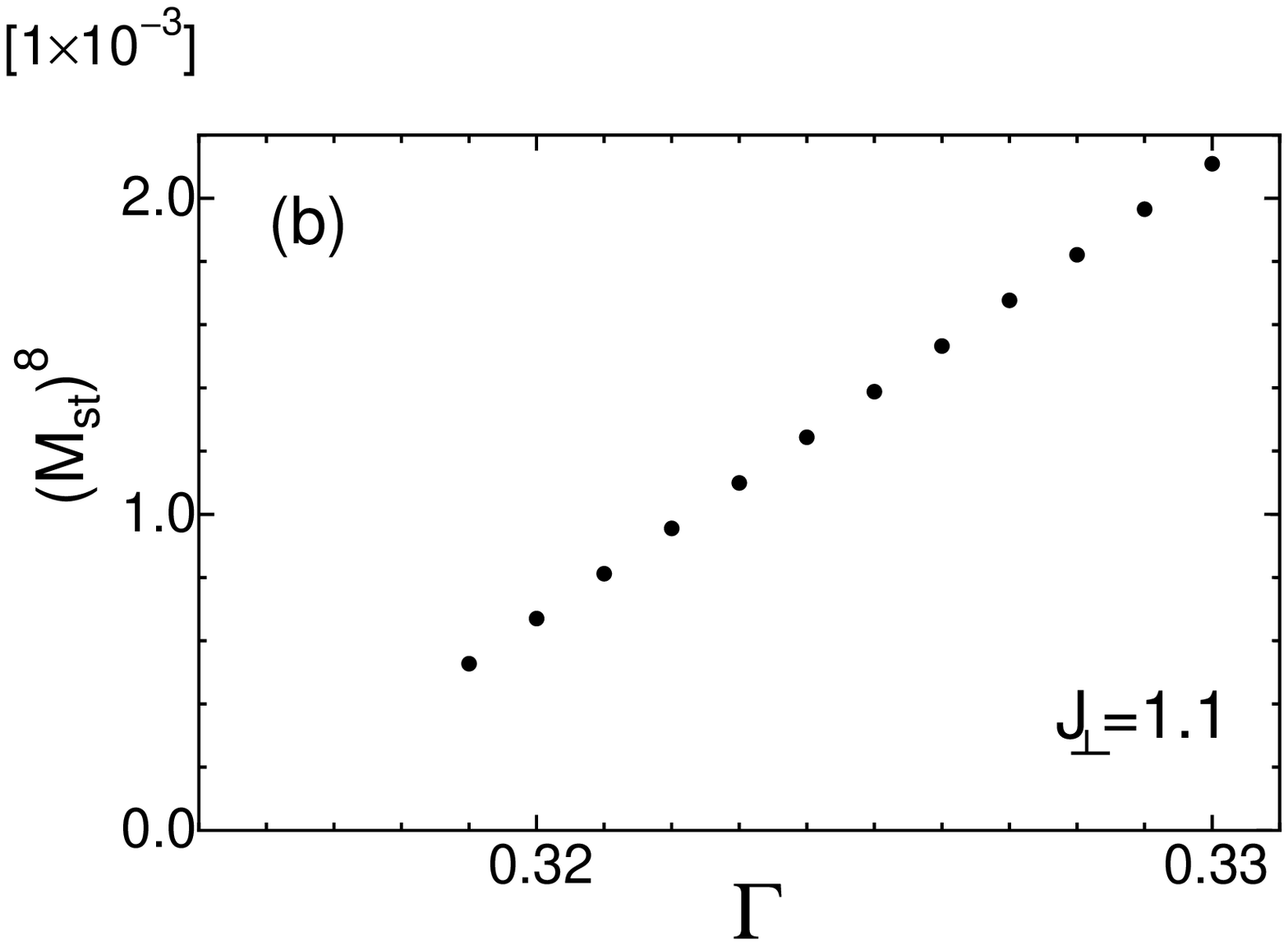}
 \end{center}
\caption{
Critical behavior of the staggered magnetizations for the  $S=1$ XXZ model: (a)$S=1$ XXZ chain with $J_\perp=1.2$  near the higher critical field $\Gamma_{c2}$  and (b) $S=1$ XXZ chain with $J_\perp=1.1$ near the lower critical field $\Gamma_{c1}$.
}
\label{bs1xxz}
\end{figure}

Let us next discuss the critical behavior at $\Gamma= \Gamma_{c1}$ and $\Gamma_{c2}$, which are also expected to belong to the 2D Ising universality class.
In Figs.\ref{bs1xxz}-(a) and (b), we show the plot of $(M_{\rm st})^8$ near the critical fields $\Gamma_{c1}$ and $\Gamma_{c2}$, respectively.
In these figures, we can recognize the good linearity.
Near the lower critical field $\Gamma_{c1}$, we have estimated the critical exponent $\beta=0.126$ for $J_\perp=1.1$.
Also near the higher critical field $\Gamma_{c2}$, we have $\beta=0.126$ and $0.127$ for $J_\perp=1.2$ and $1.4$, respectively.
These values of $\beta$  are consistent with the 2D Ising universality class.
The critical behavior of $M_x$ should correspond to that of the internal energy of the 2D Ising model, as well.
However it is difficult to estimate the exponent $\alpha$ with a direct calculation, since the expected value $\alpha=0$(log) is too weak  to be detected numerically.

\subsection{$S=1$ Heisenberg model with a single-ion anisotropy}
For the case of the $S=1$ quantum spin chains, we can consider the single-ion-type crystal field as an origin of the anisotropy, which is actually observed in the real compounds.\cite{SIexp} 
The Hamiltonian is given by
\begin{equation}
{\cal H} =
\sum_{i}
         \vec{S}_i \cdot \vec{S}_{i+1} 
        + D \sum_i (S^z_i)^2 -\Gamma \sum_{i} S^x_i.
\end{equation}
If $D<0$, the $D$-term($D \sum_i (S^z_i)^2$) stabilizes ``$\pm 1$'' spins
equivalently and hence the $D$-term is expected to make the similar
effect to the previous XXZ-type  anisotropy. 
In the following, we thus consider $D < 0$  region, where the staggered order in the $z$-direction is induced.
The phase boundary of this model is investigated by Sakai and Takahashi\cite{SaTa}, using the exact diagonalization up to 14 sites.
However the magnetization curve of this model has not been calculated yet.

\begin{figure}
 \begin{center}
  \leavevmode  \epsfxsize=65mm
  \epsfbox{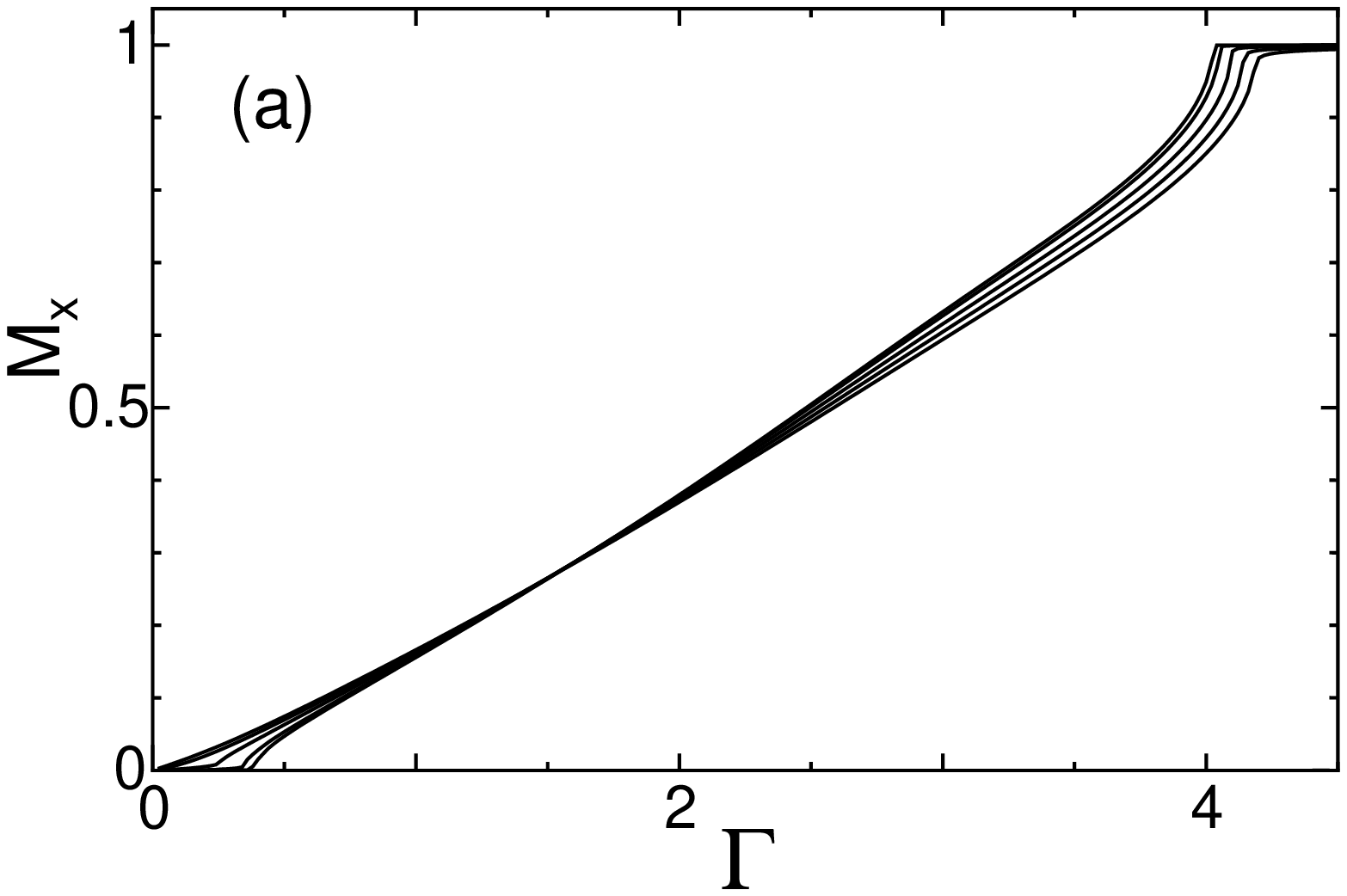}\\
  \leavevmode  \epsfxsize=65mm
  \epsfbox{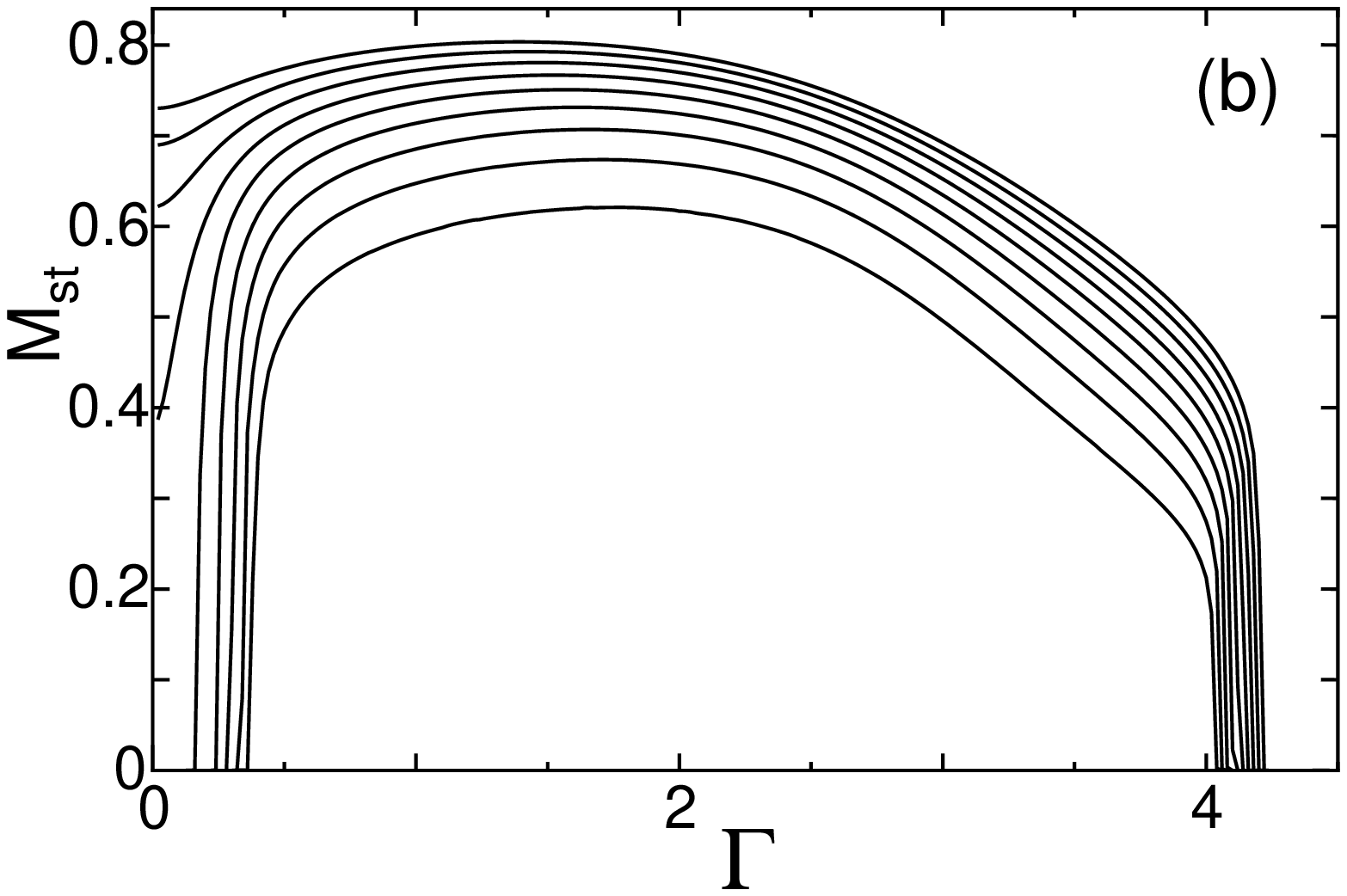}
 \end{center}
\caption{The PWFRG results for the $S=1$ Heisenberg model with the
single-ion anisotropy: (a) the uniform magnetization $M_x$ and (b) the
staggered magnetization $M_{\rm st}$. 
In Fig. (a), the curves correspond to $D=-0.4$,  $-0.3$, $-0.2$,
$-0.1$, and  $-0.05$ from left to right. 
In Fig. (b), the curves correspond to $D=-0.45$, $-0.4$, $-0.35$,  $-0.3$,
$-0.25$, $-0.2$, $-0.15$, $-0.1$, and  $-0.05$
from top to bottom.
} 
\label{figs1si}
\end{figure}

In Fig.\ref{figs1si}, we show the uniform magnetization $M_x$ and the
staggered magnetization $M_{\rm st}$.  
We can see that they exhibit the similar behavior to that of the
$S=1$ XXZ models. 
For $-0.3<D<0$, the system has the  staggered order between the two critical fields $\Gamma_{c1}$ and  $\Gamma_{c2}$. 
In $\Gamma<\Gamma_{c1}$, the system is in the Haldane phase, but  the uniform magnetization $M_x$  has a finite value due to the lack of the total-$S^z$ conservation law.
As $D$ is decreased, $\Gamma_{c1}$ becomes small,  and finally vanishes at the critical value of the anisotropy $D_c\simeq -0.3$, which is consistent with $D_c=-0.29$ calculated by Sakai et al.\cite{SaTa} 
As $D$ is decreased further,  the staggered magnetization appears at the zero transverse field, and then the reentrant behavior of the staggered magnetization can be seen. 

\begin{figure}
 \begin{center}
  \leavevmode  \epsfxsize=65mm
  \epsfbox{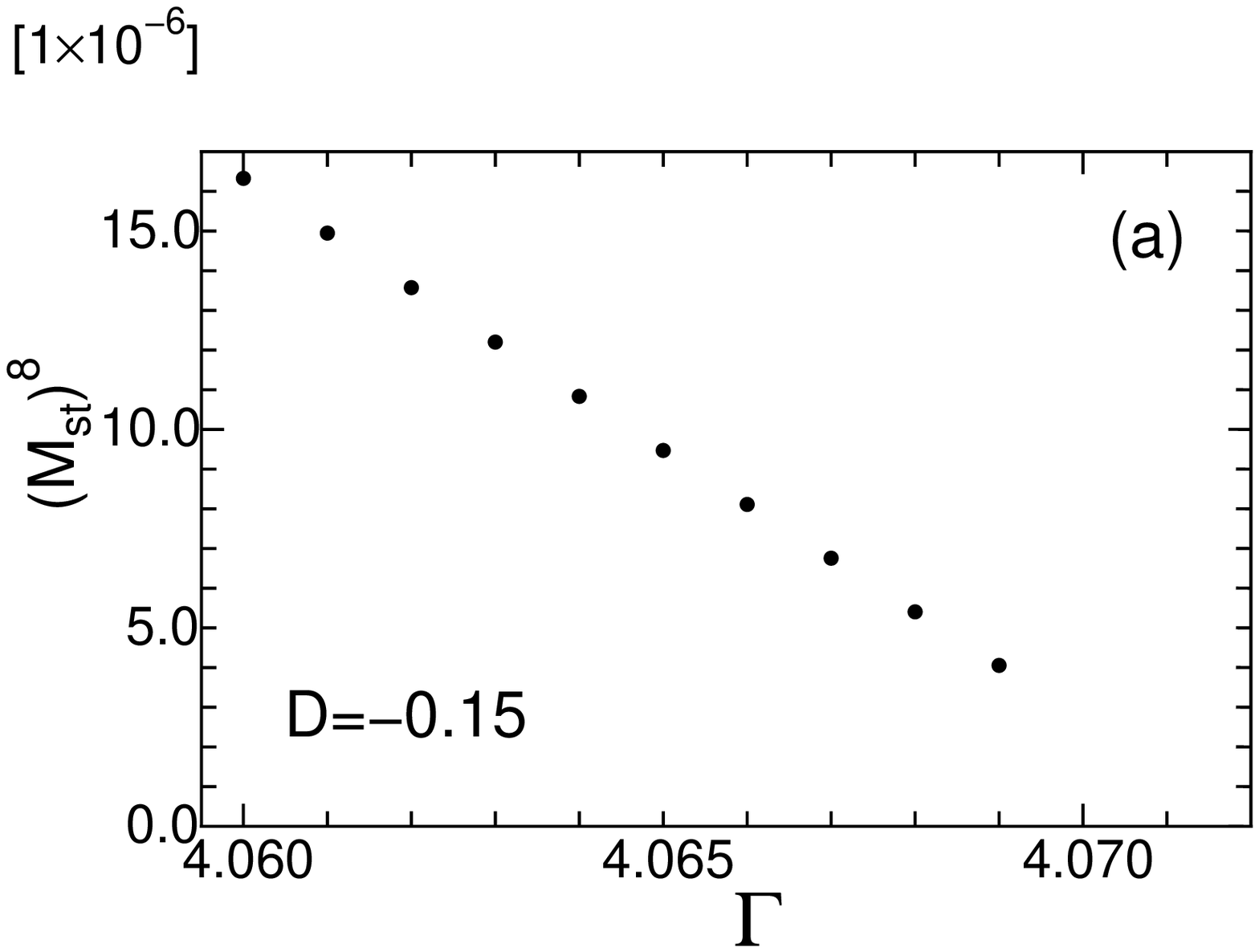}\\
  \leavevmode  \epsfxsize=65mm
  \epsfbox{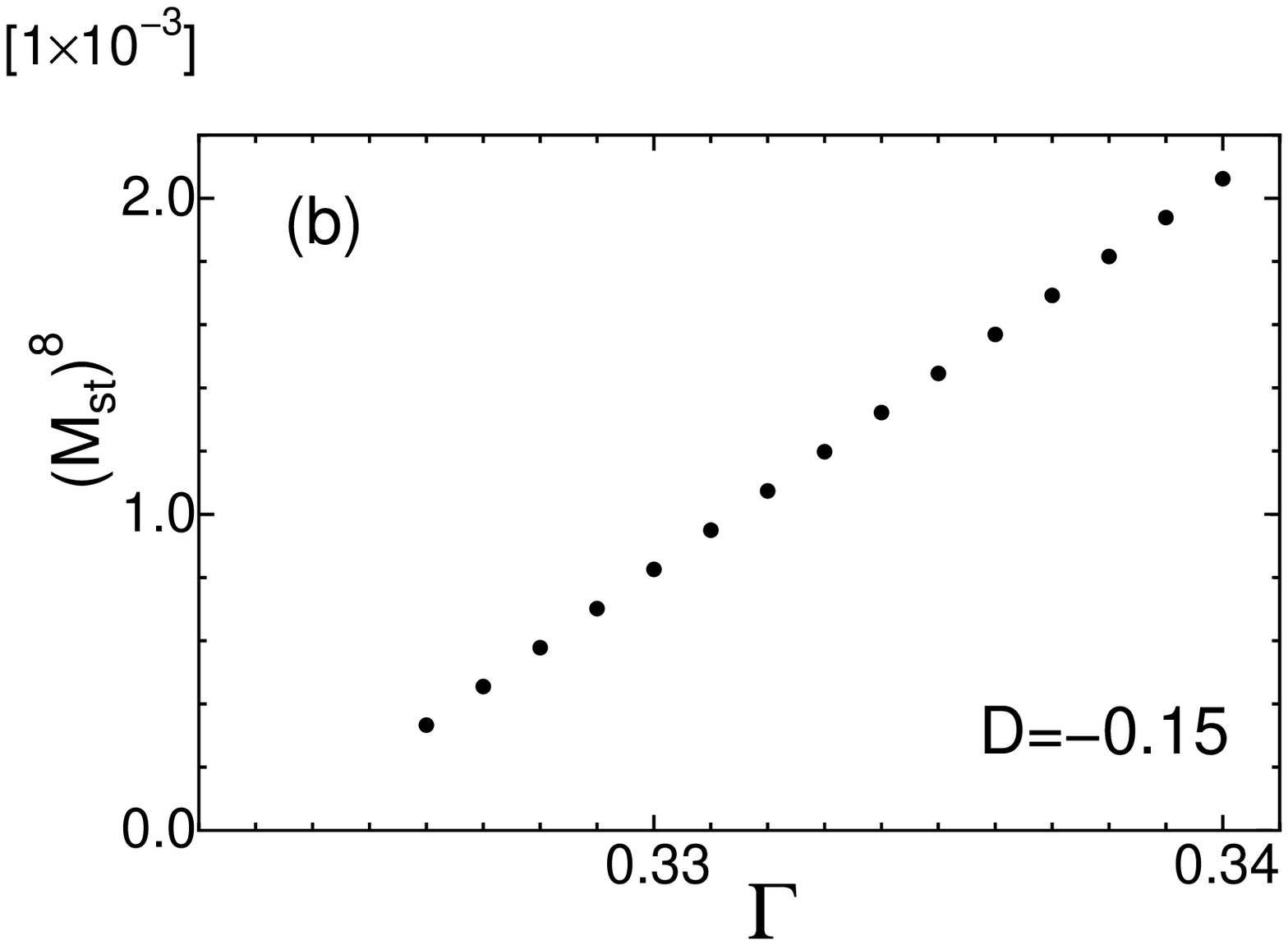}
 \end{center}
\caption{Critical behavior of the staggered magnetizations for $S=1$ Heisenberg model with the single-ion anisotropy of $D=-0.15$: (a)near the higher critical field $\Gamma_{c2}$ and (b) near the lower critical field $\Gamma_{c1}$. }
\label{figs1sibeta}
\end{figure}

Next, we plot $(M_{\rm st})^8$ near the critical fields in Fig. \ref{figs1sibeta}, where the good linearity can be seen.
For example, the estimated critical exponent at the lower-critical field is $\beta= 0.127$ for $D=-0.15$, and the ones at the higher critical fields are $\beta=0.127$ and $0.126$ for $D=-0.15$ and $-0.45$ respectively.
These values are also  consistent with the universality class of the 2D Ising model.

\section{Summary}

In this paper, we have considered the one-dimensional quantum spin chains with the easy-axis anisotropy in the transverse magnetic field. 
We have calculated the staggered magnetization $M_{\rm st}$ in the $z$ direction and the uniform magnetization $M_x$ along the $x$ direction precisely, using  a variant of the DMRG(the PWFRG method).
For the $S=1/2$ XXZ chain, we have shown that the $M_{\rm st}$ exhibits the reentrant behavior when $J_\parallel \ge 0.3984$, where the competition between the classical N{\' e}el order and the quantum fluctuations from the transverse field  and the $XY$-term plays an important role.
We have also discussed the critical behavior associated with the
staggered magnetization and verified that the universality class of
the model is the two-dimensional(2D) Ising one.  

For the $S=1$ XXZ model and the $S=1$ Heisenberg model with a single-ion anisotropy, we have shown that, if the anisotropies are not big($J_{\perp}<1.18$ or $D>-0.30$), the systems have the staggered order between  the lower critical field  associated with the Haldane gap($\Gamma_{c1}$) and the higher one connected to the saturation field of the isotropic case($\Gamma_{c2}$). 
When the anisotropies are big, the staggered order appears at the zero transverse field with the reentrant behavior, which is similar to that of the $S=1/2$ XXZ model.
We have also investigated the critical behavior at $\Gamma_{c1}$ and $\Gamma_{c2}$, by estimating the exponent $\beta$.
As a result, we have verified that they are consistent with the class
of the 2D classical Ising model.

In the actual one-dimensional compounds, various types of the anisotropy  can be considered. When analyzing the magnetization processes of such matters,  we believe that  the present results are useful.


\section*{Acknowledgments}
One of authors (Y.H.) would like to thank T. Nishino for fruitful discussions.
Y.H was supported by the Research Fellowships of the Japan Society for 
the Promotion of Science for Young Scientists.

\end{document}